\documentclass[prb,twocolumn,aps,superscriptaddress,showpacs]{revtex4}
\usepackage{amsmath}
\usepackage{graphicx}
\newcommand{\beq}{\begin{equation}}
\newcommand{\eeq}{\end{equation}}
\newcommand{\be}{\begin{equation}}
\newcommand{\ee}{\end{equation}}
\newcommand{\bea}{\begin{eqnarray}}
\newcommand{\eea}{\end{eqnarray}}
\newcommand{\barr}{\begin{array}}
\newcommand{\earr}{\end{array}}

\begin{document}

\title{Inelastic hard-rods in a periodic potential}

\author{Fabio Cecconi}
\affiliation{INFM Center for Statistical Mechanics and Complexity, 
P.le A.~Moro 2, 00185 Rome, Italy}
\affiliation{Dipartimento di Fisica, Universit\`a La Sapienza,  
P.le A.~Moro 2, 00185 Rome, Italy}

\author{Umberto Marini Bettolo Marconi}
\affiliation{Dipartimento di Fisica, Universit\`a di Camerino,
Via Madonna delle Carceri, 62032 , Camerino, Italy and
INFM, Unit\`a di Camerino}

\author{Fabiana Diotallevi}
\affiliation{Dipartimento di Fisica, Universit\`a La Sapienza,
P.le A.~Moro 2, 00185 Rome, Italy}

\author{Andrea Puglisi}
\affiliation{INFM Center for Statistical Mechanics and Complexity, 
P.le A.~Moro 2, 00185 Rome, Italy}
\affiliation{Dipartimento di Fisica, Universit\`a La Sapienza,
P.le A.~Moro 2, 00185 Rome, Italy}


\begin{abstract}
A simple model of inelastic hard-rods subject to a one-dimensional
array of identical wells is introduced. The energy loss
due to inelastic collisions is balanced by the work supplied
by an external stochastic heat-bath. We explore the effect of the
spatial non uniformity on the steady states of the system.
The spatial variations of the
density, granular temperature and pressure
induced by the gradient of the external potential are investigated
and compared with the analogous variations in an elastic system.
Finally, we study the clustering process by considering
the relaxation of the system starting from a uniform homogeneous state.
\end{abstract}
\maketitle

\section{Introduction}
Recently granular gases, i.e. a large number of macroscopic particles
colliding with one another and losing a little energy at each collision,
have been studied employing methods of statistical 
physics.\cite{general1,general2,general3,general4,zanetti}
The main practical motivation to improve the 
understanding of the dynamical and static properties
of these materials stems from their relevance 
in a variety of technological and industrial processes.\cite{Book}
Granular materials also represent a new paradigm 
in the context of non equilibrium statistical mechanics.
In fact, they are open systems which can reach a non equilibrium
stationary state when the energy loss,
determined by the dissipative interactions, 
is balanced by the work supplied by an external driving force.

A current problem in this field, motivated by a series
of recent experiments and theoretical work, 
concerns the behavior of driven granular gases 
subject to spatially non uniform 
conditions.\cite{swiss,Eggers,Lohse0,Lohse1,Lohse2,Brey,Barrat,Biwell,
Bettolo,Conti}
These conditions can be realized experimentally by shaking the gas in a
container divided in a series of identical compartments connected by holes.  
The following phenomenology is observed: for
strong shaking intensity the grains jump
from one compartment to the other, 
whereas for weak shaking intensity, they tend to break the discrete
symmetry of the system and fill preferentially only some compartments.
Such a clustering occurs because the grains, 
unlike an ideal gas of molecules, can dissipate heat
when colliding. Therefore 
a spontaneous increase of the number of particles
within a given compartment entails more collisions and  
a consequent higher energy loss. Thus the capability for the particles
to leave the compartment is strongly inhibited 
leading eventually to clustering.

In the present paper we consider a simple modeling of the multi-compartment
experiment, namely a one dimensional system of inelastic 
hard rods~\cite{sela,mackintosh,mcnamara,kadanoff}
driven by a stochastic thermostat and subject to an external
time-independent periodic potential. 
As shown in our recent paper,\cite{CDBP} the one dimensional geometry
of the model not only minimizes the amount of computational effort, 
but also allows
performing analytical guesses concerning the relevant
observables. 

As a premise, we recall that
in the case of elastic hard-rods coupled to a stochastic heat bath
and in the presence of a periodic external potential most of the equilibrium
properties can be computed by using of an exact functional relation,
proved by Percus,\cite{Percus} between an arbitrary external
potential and the equilibrium average density profile. 
It is natural to ask how the behavior of the elastic
system will be affected
upon switching on
the inelasticity of the collisions.  The equilibrium state
of the elastic hard-rod system is replaced by a corresponding 
non equilibrium steady state of the inelastic system.
The comparison between the properties of the two systems offers
a useful procedure to understand the interplay between
the inelasticity, the excluded volume effect and 
the external potential.
As we shall see below
the model shows a variety of non equilibrium states according to the
values of the control parameters,  such as the external potential,
the density and the driving force.
These states range from spatially periodic configurations
to non periodic configurations in agreement with the 
experimental findings.

The paper is organized as follows. In section II, we introduce the model
we employed to study the effects of compartments in granular gases and we 
briefly discuss the main feature of the dynamics. In section III, we 
compare the steady state properties of the model against
the corresponding properties of the equilibrium elastic system. 
In section IV, we consider the relaxation and clustering properties of the 
system. Finally, conclusions are presented in section V.

\section{A model of driven granular gas in compartments}
The main issue of the present paper is to study 
how the properties of a
granular fluidized system are affected by the presence of a static
external non uniform potential.  We will show that such a
confining potential can enhance the tendency toward cluster formation,
generating inhomogeneities similar to those observed in some real
granular systems placed in compartments. 
The model studied
consists of a set of $N$ impenetrable particles of equal masses, $m$, and
lengths, $\sigma$, moving on a ring under the influence
of the periodic potential $V(x)$. 
Two mechanisms control the energy of the particles: the interaction with an
heat bath and the collisions among the particles. 
The collisions are instantaneous and binary and dissipate
a fraction $(1-r^2)/2$  of the total kinetic 
energy of the pair ($r$ being the coefficient of restitution). 
The heat bath, instead, supplies  an amount
of energy which prevents the system to
come at rest. In this paper we employ, a stochastic energy source, namely
the thermostat previously employed
in Refs.~[\onlinecite{mackintosh,Swift,Biwell,CDBP}].  
The coupling of the particles to
the heat-bath consists of the combination of a velocity dependent frictional
force and a random kick. 
We first write the evolution equation for the generic particle $i$ in the 
system when the collisions are not taken into account:
\begin{equation}
m \frac{d^2 x_i}{dt^2}=-m \gamma\frac{d x_i}{dt} - 
V'(x_i)+\xi_i(t)
\label{kramers}
\end{equation}
where, $x_i$ ($i=1,N$), $\gamma$ and $V'(x)$ indicate 
the position of particle $i$,
the friction coefficient and the spatial derivative of the
external static potential, respectively.
The kicks, $\xi_i(t)$, are distributed according to a
Gaussian law characterized by  
$$
\langle\xi_i(t)\rangle=0
$$
and 
$$
\langle \xi_i(t)\xi_j(s) \rangle = 2 m \gamma
T_b \delta_{ij} \delta(t-s)
$$
The temperature $T_b$ determines the intensity of the heat bath. 

\begin{figure}[htbp]
\includegraphics[clip=true,keepaspectratio,width=8.cm]{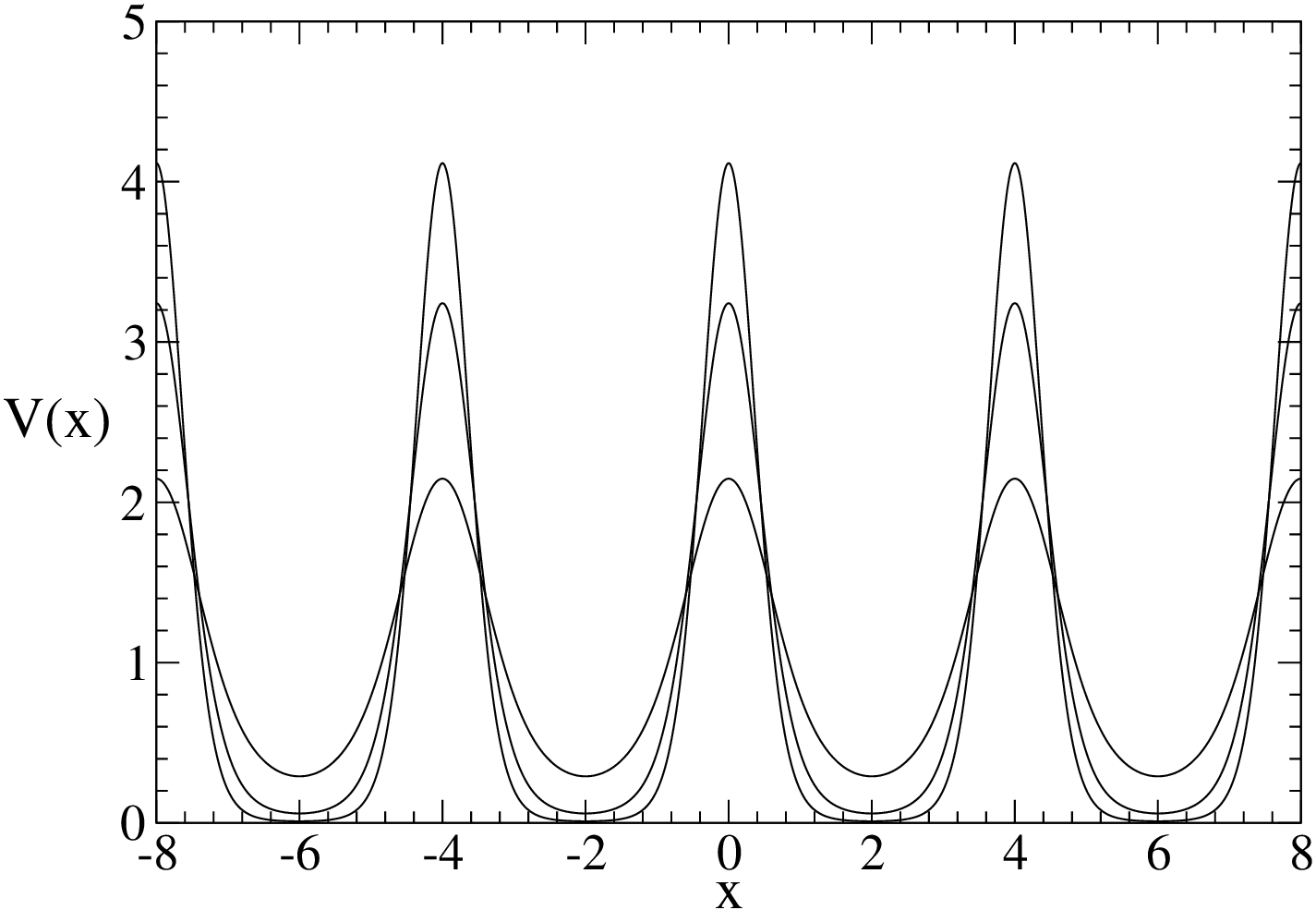}
\caption
{Behavior of $V(x)$ for $\alpha=1,2,3$, as $\alpha$ increases,
the barriers approximate the condition of walls and the wells become more
square-like.}
\label{fig_poten}
\end{figure}

The external potential has a spatial period $L/M$, where $L$ is
the system size and $M$ is the number of potential wells
(i.e. number of local minima) in the interval $[-L/2,L/2]$.
Each well mimics, in our
simulation,  a compartment in Lohse's
experiments.\cite{Lohse0} The boundary conditions are assumed
to be periodic.  We found that a rather convenient potential to
use in numerical experiment is that of
Refs.~[\onlinecite{Magnasco,Das}]:
\begin{equation}
V(x)=\frac{\exp\{\alpha \cos(2\pi x M/L)\}}{I_0(\alpha)}\;,
\label{eq:pot1}
\end{equation}
where the zeroth-order Modified Bessel Function,\cite{Abramowitz} 
$I_0$, in the denominator, enforces
the normalization condition $1/L\int_0^{L} dx V(x) = 1$.  
The shape of the potential $V(x)$ depends upon the
parameter $\alpha$ (Fig.~\ref{fig_poten}): as it increases, the shape
of the barriers become more square-like, approximating the condition of
walls.

Collisions redistribute the energy and the momenta among the particles.
For a nearest neighbor pair a collision event takes place when
the separation  $x_{i+1}-x_i$ equals their length, $\sigma$
and determines a change  of the velocities of the
two particles according to the rule:
\begin{equation}
v_i' =  v_i - \frac{1+r}{2}(v_i - v_j).
\label{collision}
\end{equation}
where the prime indicates post-collisional variables.
Notice that in the elastic case, $r=1$, the collisions do not change
the total energy and that the average kinetic energy  per particle
attains asymptotically the value $T_b$. This state of affairs is
altered for values of the coefficient of restitution less than
one and the average kinetic energy per particle reaches (for $V(x)=0$)
a value smaller than $T_b$.

In the following we will assume a unitary mass $m=1$.

The basic phenomenology of the inelastic model can be understood 
by considering the two competing effects
controlling the dynamics: the hard core
interaction and the collisional dissipation.  The former favors the
equidistribution of the particles in the various wells, whereas the
latter promotes the formation of clusters in some wells, thus breaking
the periodic symmetry of the system.  If the particles were
independent, the passage from one well to the next would occur via an
activated jump process, characterized by a typical exit time, $\tau$,
proportional to $\exp(\Delta V/T_b)$, where 
$\Delta V = 2\sinh(\alpha)/I_0(\alpha)$ represents the energy barrier 
between adjacent wells. 
In other words, a local density excess of particles would be
washed out by a diffusive process whose diffusion constant, $D$, has
the Arrhenius dependence, $D=D_0\exp(-\Delta V/T_b) $, with
$D_0 = T_b/m\gamma$.
 
In a recent paper~\cite{Biwell} we have shown that, in a granular toy
model consisting of two wells and only two particles, two different
jump processes can be clearly identified: the free-jump, which occurs
when the escape is from a singly occupied well, and the
correlated-jump that occurs when the escape involves a doubly occupied
well. The resulting scenario is that the jumping dynamics remains
still Arrhenius-like but the jump-rate strongly depends on the
occupation of the well from which the jump occurs, i.e. 
\begin{equation} 
\tau_k \propto \exp\bigg(\frac{\delta V_k}{T_k}\bigg)
\label{newkramer}
\end{equation}
where $k$ indicates the occupation number ($k= 1$ or $2$).
In the multi-well and many particle systems, we shall discuss hereafter,  
the phenomenon is enhanced and the escape rate for a given well 
will  depend on the number of particle inside that well. 

The steady state properties of our model are investigated by means of
molecular dynamics simulations of Eq.~(\ref{kramers}), for an ensemble
of $N = 1024$ hard particles of length $\sigma=0.2$, in a box of
size $L=1024$. The heat bath parameters are 
$\gamma=0.2$ and $T_b = 1.0$.  The potential
constants $\alpha=1$ and $M=100$, imply $100$ wells (compartments) of
width $w=L/M=10.24$, separated by an energy
barrier of height $\Delta V = 1.856$.  The computer 
implementation of the granular dynamics has been realized by a suitable 
modification of the event-driven algorithm to take into account the effects 
of the potential and of the heat-bath.  Particle positions and
velocities within two consecutive collisions are updated according to
a second order discretization scheme for the
dynamics Eq.~(\ref{kramers}). 

Simulations start from initial conditions corresponding to equally spaced
non overlapping particles.
After a transient time $t_0$, the system reaches a
steady state characterized by robust statistical properties
and one measures the 
generic observable, $A(t)$, by performing the time average
\begin{equation}
\langle A \rangle = \frac{1}{t-t_0} \int_{t_0}^t d\tau A(\tau)
\end{equation}
over an appropriate interval, $t-t_0$.

\section{Properties of the steady state}
We analyze first the reference elastic system ($r=1$)
for which a true equilibrium state exists. Most of its properties
are exactly known. For instance,
the grand-potential $\Omega[\rho(x)]$, from which
all the equilibrium properties can be deduced,
is a functional of the density $\rho(x)$ whose explicit form
reads
\begin{eqnarray*}
\Omega[\rho] = T_b \int^{L/2}_{-L/2} dx 
\rho(x)\bigg[\ln\bigg\{\frac{\rho(x)}{1 - \eta(x)}\biggr\} - 1 + 
\frac{V(x) -\mu}{T_b}\bigg]
\end{eqnarray*} 
with $\mu$ being the bulk chemical potential of the system and 
$$
\eta(x) = \int_{x-\sigma}^x ds \rho(s)
$$ 
the local packing fraction.
Extremising 
$\Omega[\rho]$ with respect to $\rho(x)$ yields the following 
Euler-Lagrange equation
\begin{equation}
\ln \frac{\rho(x)}{1-\eta(x)} = \beta[\mu - V(x)] -
\int^{x+\sigma}_x dy \frac{\rho(y)}{1 - \eta(y)}
\label{euler}
\end{equation}
whose solution yields the equilibrium density profile.
Notice that inserting this solution into $\Omega$ 
one obtains the value thermodynamic grand potential 
of the system~\cite{Dieterich}
\begin{equation}
{\tilde \Omega} =
-T_b \int^{L/2}_{-L/2} dx \frac{\rho(x-\sigma)}{1-\eta(x)} 
\label{ome}
\end{equation}
Using Eq.~(\ref{ome}) in the uniform limit ($V(x)=0$), we obtain 
the bulk pressure
$-{\tilde \Omega}/L = P =T_b \rho/(1-\rho)$.

\begin{figure}[htb]
\includegraphics[clip=true,width=8.cm,keepaspectratio]{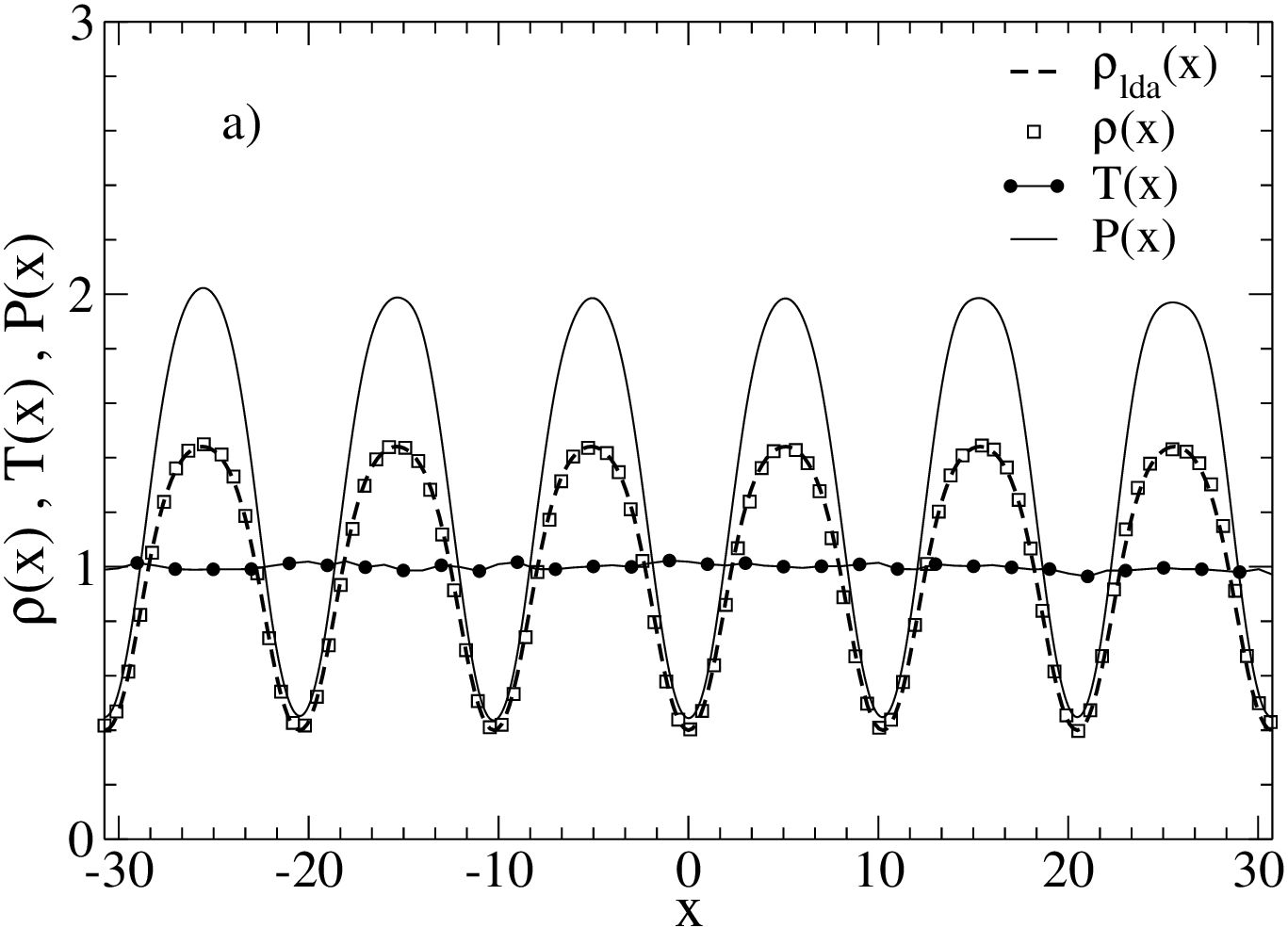}\\
\includegraphics[clip=true,width=8.cm,keepaspectratio]{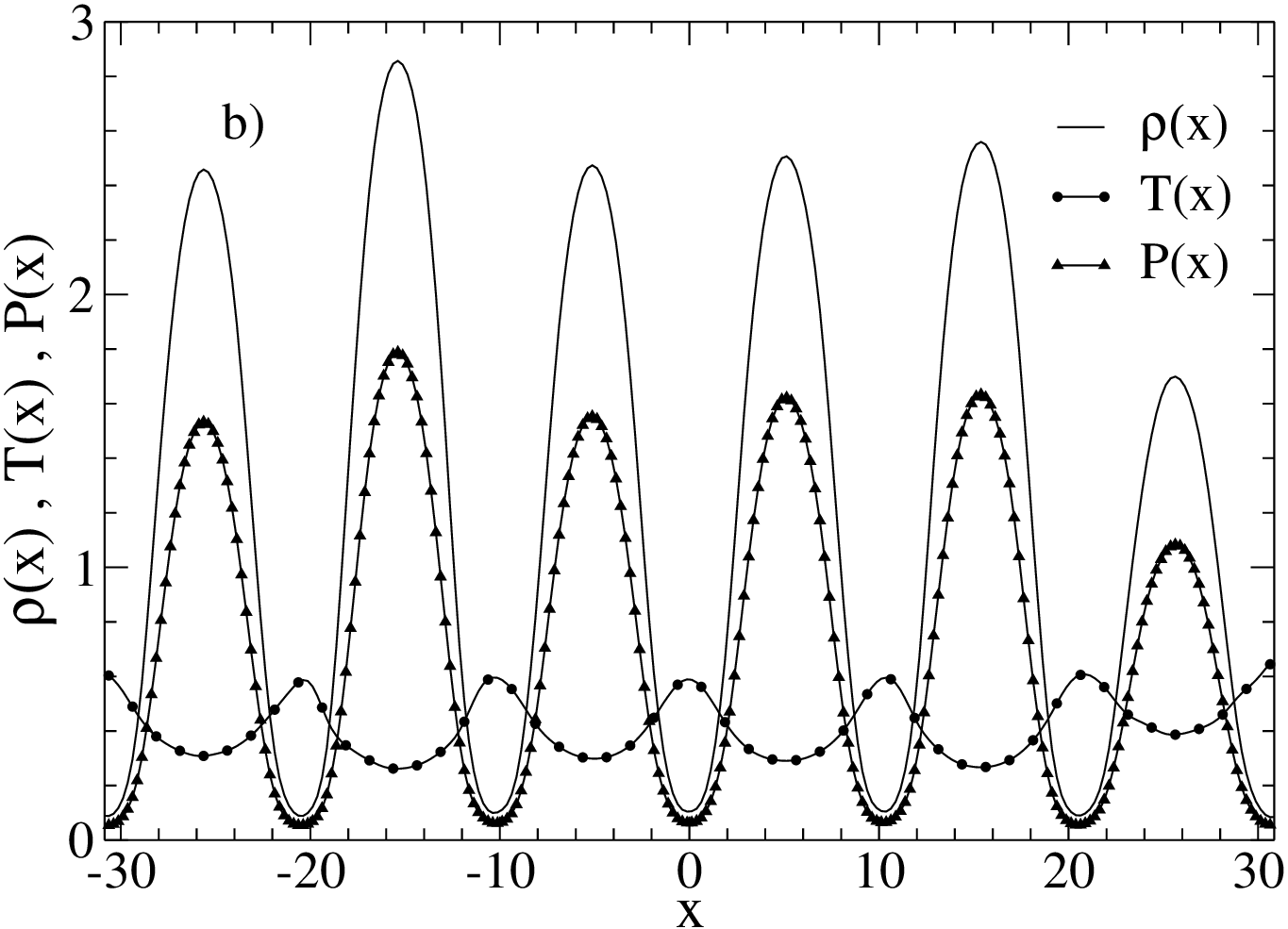}\\
\includegraphics[clip=true,width=8.cm,keepaspectratio]{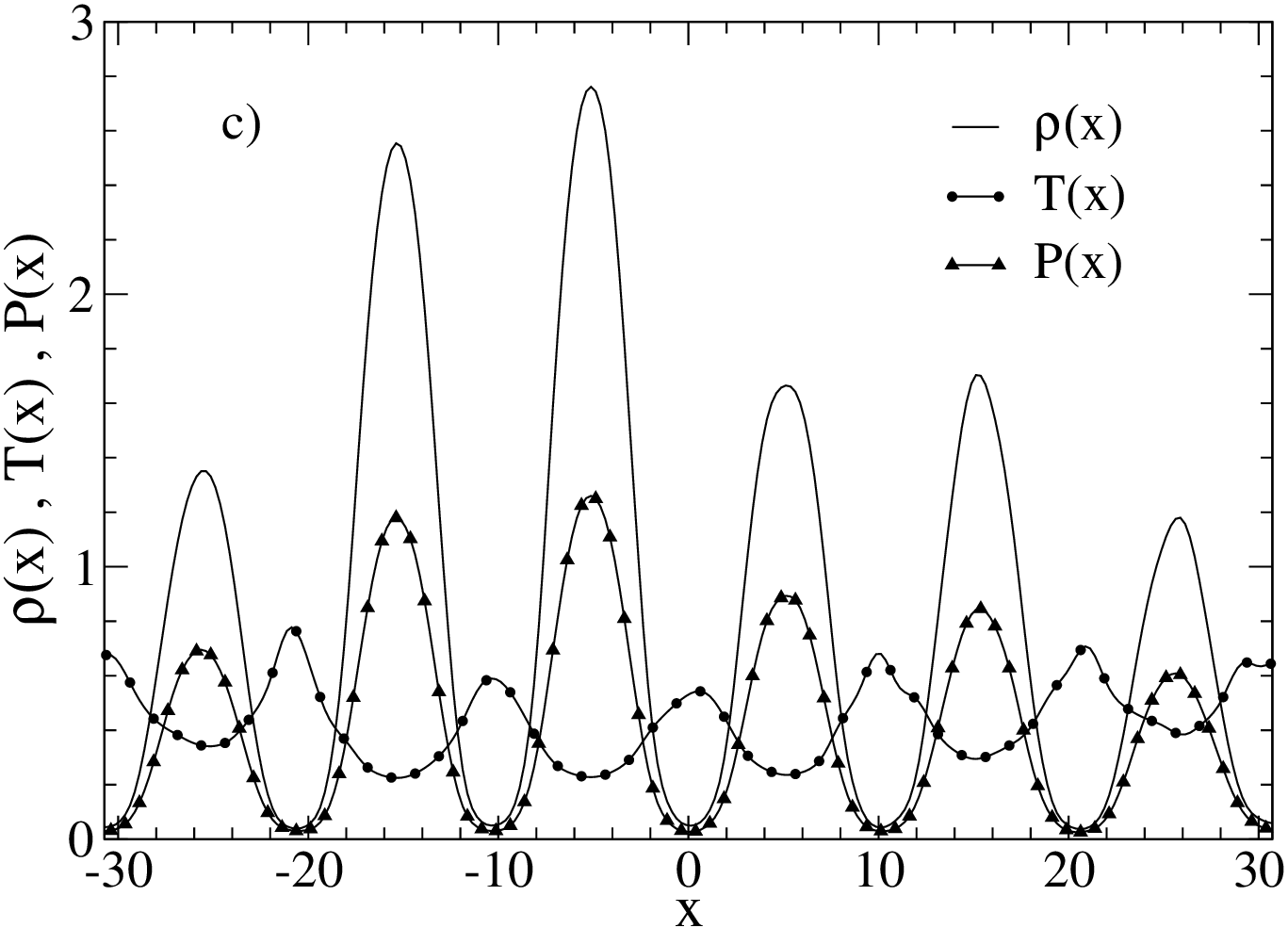}\\
\caption{\label{fig:profiles}
Profiles of density, temperature and pressure
in a system of $N=1024$ rods, size $L = 1024$, diameter $\sigma=0.2$, $M=100$ 
potential wells and  $r=1.0, 0.8, 0.6$ (a,b,c respectively). 
the granular temperature varies with the position 
and is anti-correlated with
the density profile.}
\end{figure}

Equation~\eqref{euler} can also be used to test the 
average profile obtained from the numerical solution
of the dynamical equations.
Since the potential $V(x)$ varies slowly over the
hard-core diameter $\sigma$ (i.e. $L/M>>\sigma$), the Euler-Lagrange
equation can be solved in the local density approximation (LDA)
\begin{equation}
V(x) + T_b \bigg\{\ln\bigg[ \frac{\rho(x)}{1 - \sigma\rho(x)} \bigg]
+ \frac{\sigma\rho(x)}{1-\sigma\rho(x)}\bigg\} = \mu.
\label{eq:LDA}
\end{equation}
The density profile $\rho_{LDA}(x)$ 
solving  Eq.~(\ref{eq:LDA}), is shown along with
the corresponding quantity obtained from the numerical simulation, 
in figure~\ref{fig:profiles}a  and an excellent 
agreement is found. In the non uniform case, the pressure 
is obtained from 
the hydrostatic equilibrium equation: 
\begin{equation}
\frac{d P(x)}{dx}+\rho(x) V'(x)=0
\end{equation}
 
As we shall see later the grand potential functional
is related  
to the probability distribution of  numbers of particles occupying
each cell.

In the elastic system the temperature and the pressure can 
be obtained in two equivalent ways: the first is through their kinetic
definition the second via their thermodynamic definition.
On the contrary in the inelastic system 
both quantities can be obtained only via their mechanical definitions.
Then temperature is computed through the kinetic energy
$$
T_g = m \sum_{i=1}^{N} \langle v_i^2 \rangle /N. 
$$
The pressure is evaluated by considering
the impulse transferred across a surface in the unit of
time~\cite{Allen,Barrat} and is given by the the sum of the
ideal gas term $P_{id}=T_g\rho $ and the collisional $P_{exc}$:
\begin{equation} 
P = \rho T_g(\rho) + \frac{\sigma}{Lt_{ob}}\sum_{k=1}^{M_c} \delta p_k.
\label{eq:Ptot}
\end{equation}
Here $t_{ob}$ is the observation time, the sum runs over the $M_c$
collisions occurring during
$T_{ob}$, and $\delta p_k = m \delta v_k$ represents the impulse
variation in the $k$-th collision.  As shown in
Fig.~\ref{fig:profiles}a, the pressure profile follows the same
pattern as the density. 
Let us remark that with the parameters
here employed the results of LDA are in
fair agreement with the corresponding results of the simulations, 
thus suggesting
that even in the inelastic case a treatment of the excluded
volume effect as a local density effect might be appropriate.

Let us consider the effect of the inelasticity. In 
Figs.~\ref{fig:profiles}b and~\ref{fig:profiles}c,
obtained with $r=0.8$ and $r=0.6$,
one observes two novel features 
the temperature profiles become non uniform
and the density ceases to be periodic. In fact, the kinetic temperature 
decreases in the more populated regions, due to the higher 
collision rate. The particles being less energetic, when they belong to
a cluster, tend to remain trapped even more in the wells. 
 
Notice that the average pressure is lower than the
corresponding pressure in the elastic system
at the same average density, due to the lower value of $T_g$ with
respect to $T_b$.  It also varies from well to well, in
apparent contradiction with the law of hydrostatic equilibrium.  In
fact, the state described by Figs.~\ref{fig:profiles}b
and~\ref{fig:profiles}c is not stationary, 
but slowly evolves with a characteristic rate associated to the average 
exit time from the wells, much larger than in the elastic case. 

\begin{figure}[htb]
\includegraphics[clip=true,width=8.cm,keepaspectratio]{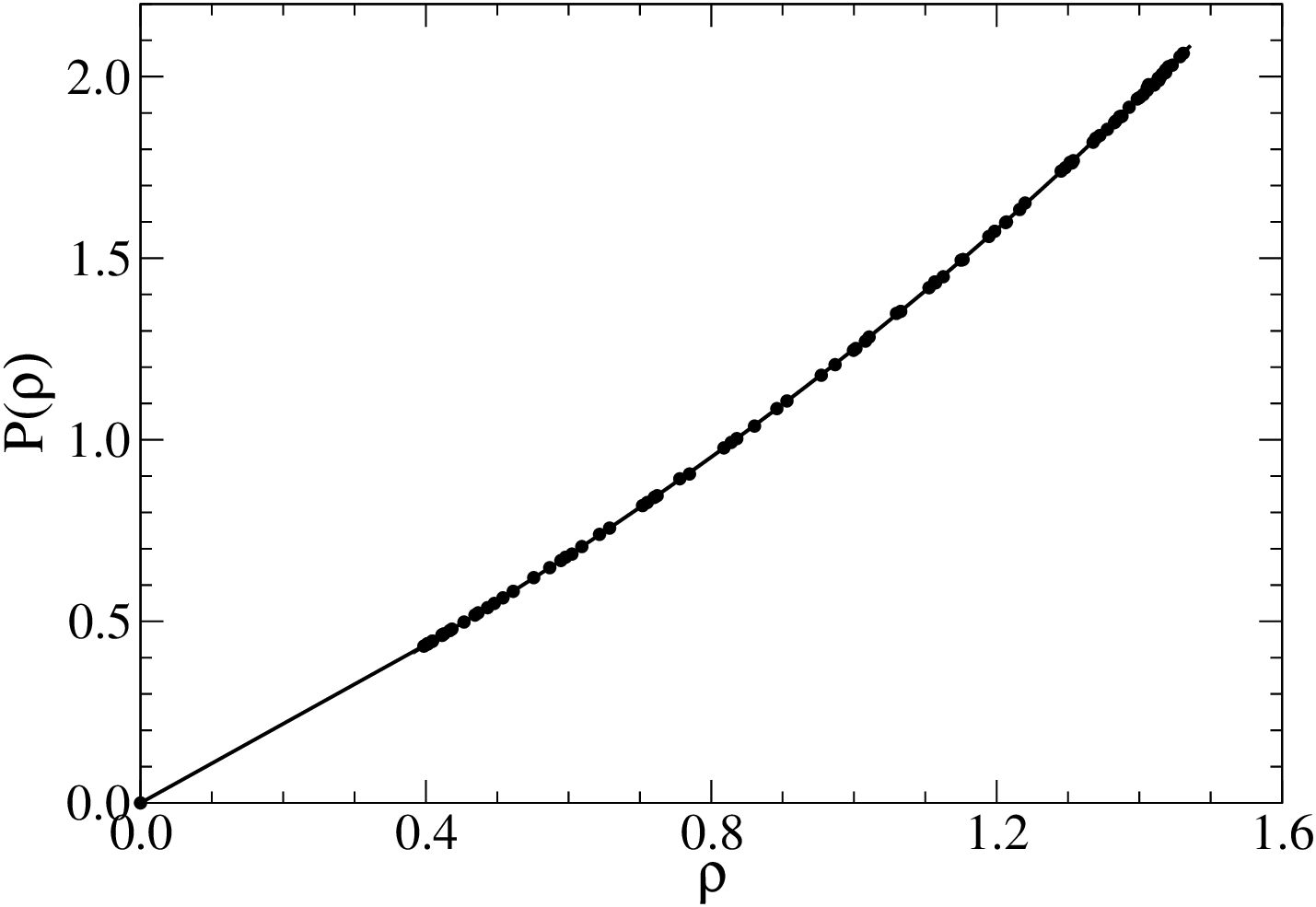}
\caption{Scatter plot of pressure vs. density for the
elastic system. Points are obtained
by plotting data of Fig.~\ref{fig:profiles}a in a
parametric plot $\{\rho(x),P(x)\}$ which eliminates the $x$-dependence.
Solid line indicates the exact result~\eqref{eq:Tonks}.}
\label{fig:pressE}
\end{figure}

At this point, it is interesting to discuss the existence of a
relationship between the quantities $\rho,P,T$, in a granular
gas, i.e. of an ``equation of state''.  
In Fig.~\ref{fig:pressE} we
plot parametrically the numerical values of the pressure versus the
density obtained by profiles in Fig.~\ref{fig:profiles}a for the
elastic system.  We notice that data are in good agreement 
with the exact Tonk's formula~\cite{Tonks}
\begin{equation}
P(\rho)= T_b \frac{\rho}{1 - \sigma \rho}
\label{eq:Tonks}
\end{equation}
  
In a recent paper\cite{CDBP} we
generalized expression~(\ref{eq:Tonks}) to the case of a granular
system with no external potential, and found:
\begin{equation}
P(\rho)= T_g(\rho) \bigg[\rho+
\frac{\sigma \rho^2}{1-\sigma \rho}\bigg]\;=\; 
T_g(\rho)\frac{\rho}{1-\sigma \rho}.
\label{eq:pressure}
\end{equation}
The equation~\eqref{eq:pressure} is equivalent to Eq.~\eqref{eq:Tonks},
upon replacing $T_b$ with $T_g(\rho)$.  
The granular temperature is a function of the density
since its value is determined by the balance between
the average power dissipated in the collisions and the
the power supplied by the thermal bath. Therefore the larger
the number of collisions (i.e.
the larger the density) the lower the granular temperature,
which reads
\begin{equation}
T_g(\rho) = \frac{T_b}{1+\frac{1-r^2}{2\gamma}\frac{\rho}{1 - \sigma\rho}
\sqrt{\frac{T_g}{m}}}.
\label{eq:Tg}
\end{equation}
Formulae \eqref{eq:Tg} and \eqref{eq:pressure} have been tested in the
homogeneous case and shown to provide  reasonable fits to the 
numerical data.\cite{CDBP}

\begin{figure}[htb]
\includegraphics[clip=true,width=8.cm,keepaspectratio]{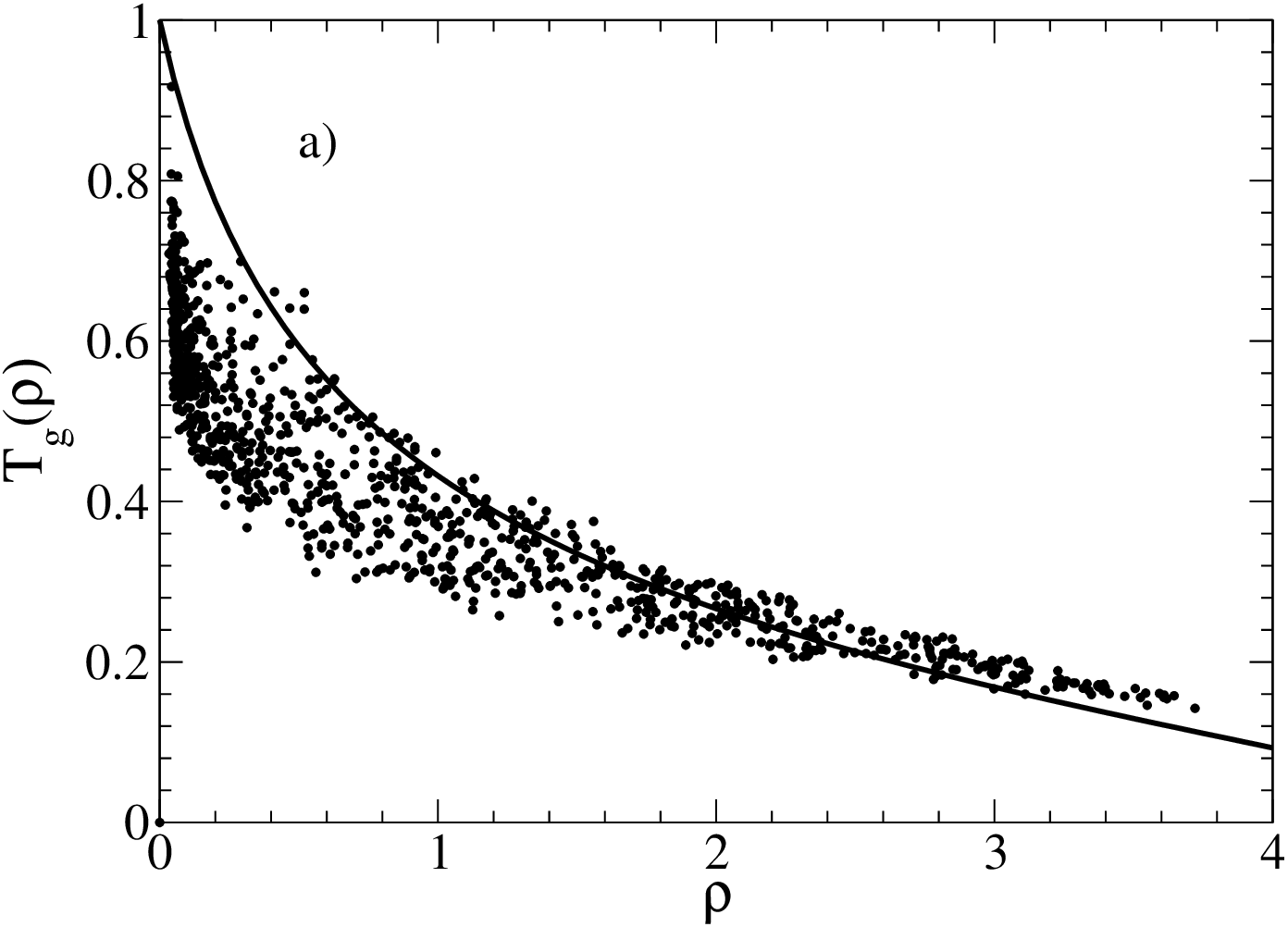}
\includegraphics[clip=true,width=8.cm,keepaspectratio]{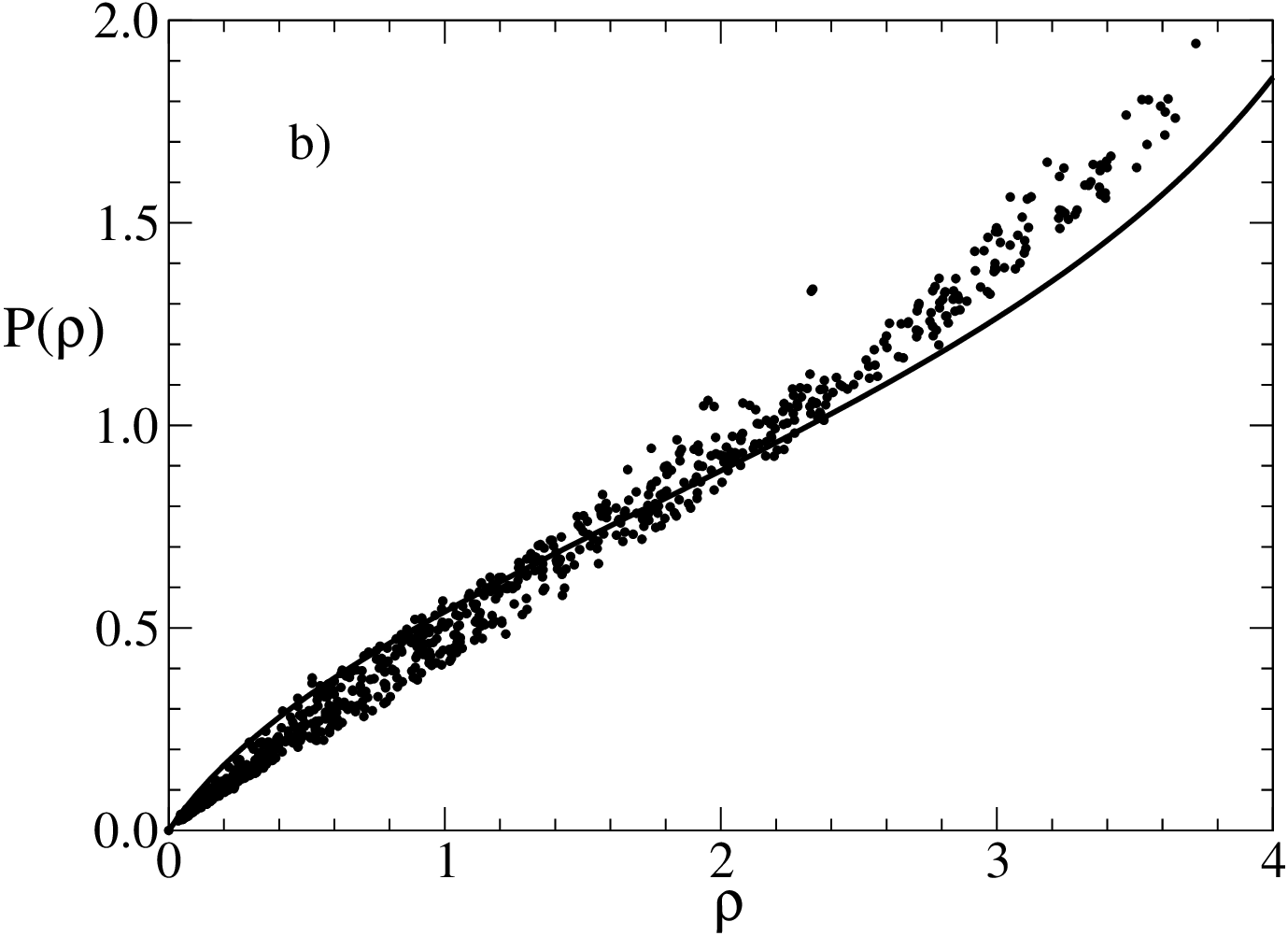}\\
\caption{\label{fig:state}
Scatter plot of temperature a) and pressure b) vs. density for a system
with an inelasticity coefficient $r=0.6$. Points are obtained from the 
profile data of Fig.~\ref{fig:profiles}c.
Solid lines indicate the theoretical estimate for temperature~\eqref{eq:Tg} 
and pressure~\eqref{eq:pressure} derived in Ref.~[\onlinecite{CDBP}] 
for the system without potential.
We notice the systematic deviation of the
measured temperature from the theoretical estimate. Its origin is probably
due to the large density fluctuations in the boundary regions 
separating the wells.}
\end{figure}

In the inhomogeneous case ($V(x)\neq 0$) with inelasticity ($r<1$), 
we extracted from the numerical simulations 
the density, temperature and pressure profiles and performed the
parametric plots shown in Figs.~\ref{fig:state}a, \ref{fig:state}b. 

One observes that the plots
deviate from the theoretical predictions and are more scattered.
The reason for the breakdown of formulae~\eqref{eq:pressure}
and \eqref{eq:Tg} can be traced back to the inaccurate
evaluation of the collision rate in the presence of the confining
potential.
The presence of the wells, indeed, strongly increases
the collision rate at the bottom of each well, thus enhancing the effects
of inelasticity.

\section{Dynamical properties}
In this section we focus our attention on the clustering process which
occurs spontaneously, starting from
a uniform configuration, when the temperature of the 
heat bath is comparable
with the energy of the potential barriers.
Since the collisional cooling determines a decrease of the local
temperature in the regions of high density, some compartments, randomly
selected by the dynamics, may act as germs for the nucleation of a
clustering process. Thus
after a long transient, the occupation of few compartments may grow
at the expense of the remaining which become less densely populated.

An immediate and effective description of the clustering phenomenon
can be achieved by analyzing the statistics of the occupation of the
wells.  We study the probability that the $i$-th well is occupied by
$n_i$ particles.  In the ideal gas, (i.e. independent and
identically distributed particles) the distribution function of $N$
particles into $M$ identical cells follows a Poisson law:
\begin{equation}
P(n_i)=\frac{\lambda^{n_i}\exp{(-\lambda)} }{n_i !}
\label{eq:poisson} 
\end{equation}
where $\lambda=N/M$ is the mean (also most probable) occupation value, and
the variance is $\sigma^2=pN=N/M = \lambda$.
It is instructive, also in the light of the discussion below, to link  
the distribution~(\ref{eq:poisson}) to the
grand potential functional. Let us consider the
probability of having a configuration with $n_1,n_2,..,n_M$ particles
in the $M$ compartments:
\begin{equation}
\Pi[\{n_s\}] = \prod_{s=1}^M P(n_s)
\label{eq:poisson2}
\end{equation}
For the equilibrium system we write 
\begin{equation}
\Pi[\{n_s\}] ={\cal N}\exp(-\Omega[\{n_s\}]/T_b)
\label{eq:prob}
\end{equation}
where ${\cal N}$ is a normalization factor and $\Omega$ indicates the
ideal-gas coarse grained grand potential
\begin{equation}
\Omega[\{n_s\}]=T_b\sum_{s=1}^M[ n_s(\ln n_s -1)-\mu n_s]
\label{eq:omega}
\end{equation}
The external potential, whose role is merely to confine
the particles into the wells, has been eliminated replacing
the original microscopic density field
$\rho(x)$ with the coarse
grained variables $n_i$.
Finally, substituting Eq.~(\ref{eq:omega}) into Eq.~(\ref{eq:prob})
we find: 
\begin{equation}
\Pi[\{n_1\}]={\cal N}\prod_{s=1}^M 
\frac{(\mbox{e}^{\mu/T_b})^{n_s}}{{n_s}!}
\end{equation}
which is identical to expression~(\ref{eq:poisson2}),  
provided $\lambda$ is replaced by  $\mbox{e}^{\beta \mu}$ and 
introducing a suitable normalization factor ${\cal N}$.

Such an ideal gas formula has to
be  modified in the case of the elastic hard-rod system,
because the hard core repulsion induces correlations among the positions
of the particles.
The excluded volume constraint  can be accounted for 
in a simple fashion, by 
replacing the ideal gas
functional, $\Omega$, by the corresponding hard-rod coarse grained 
functional:
\begin{equation}
\Omega_{hr}[\{n_s\}]=T_b\sum_{s=1}^M \bigg\{n_s[\ln n_s -1 -
\ln (N_m-n_s)]-\frac{\mu n_s}{T_b}\bigg\}
\end{equation}
where $N_m$ is the maximum number of non overlapping rods 
which can occupy a single well.
The extremization of $\Omega_{hr}$ with respect to $n_s$ yields
the link between the chemical potential $\mu$, $N_m$ and the mean 
occupation $\lambda$ 
\begin{equation}
\mu = T_b\bigg[\ln \lambda -
\ln (N_m-\lambda)+\frac{\lambda}{N_m-\lambda}\bigg].
\end{equation}

\begin{figure}[htbp]
\includegraphics[clip=true,keepaspectratio,width=8cm]{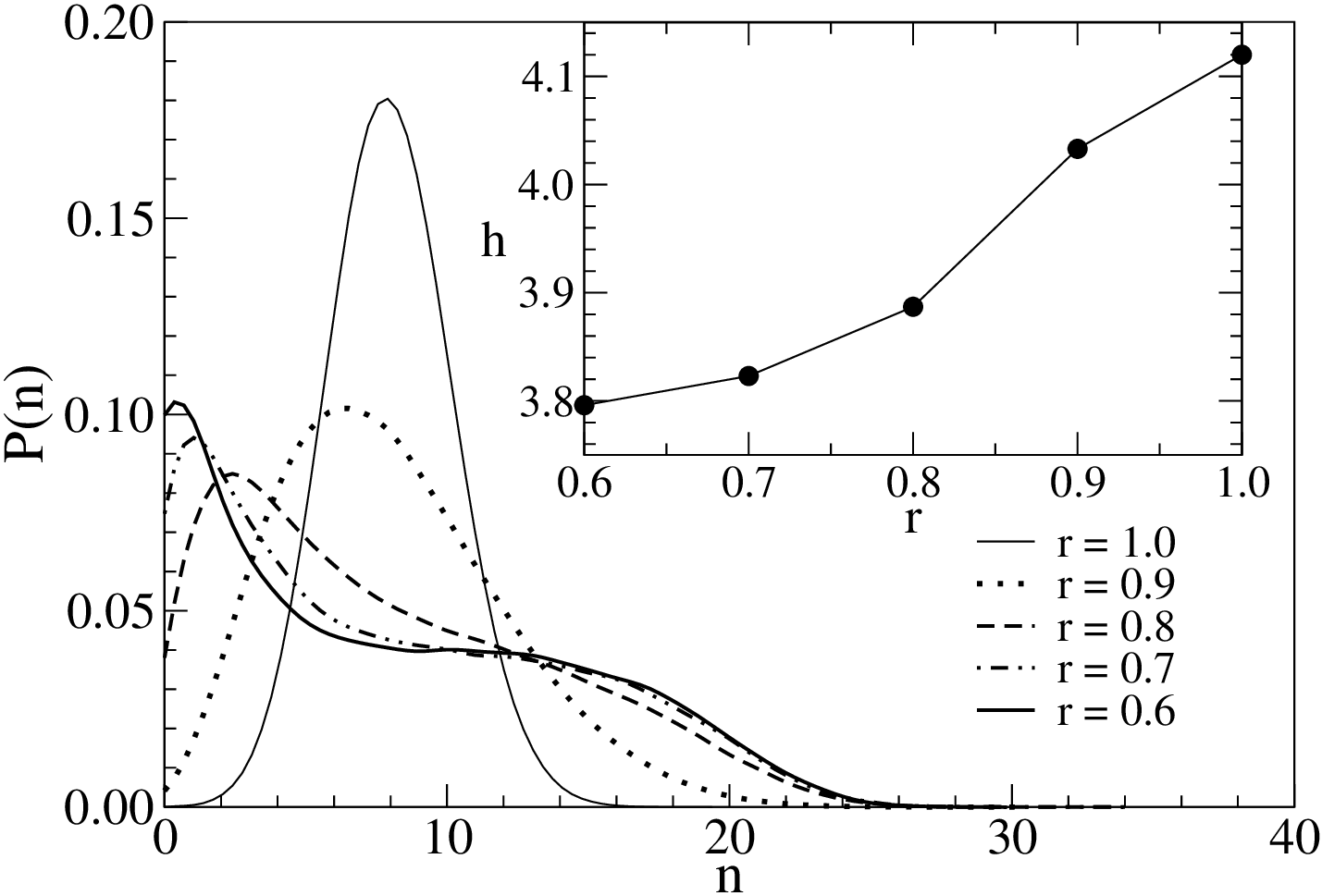}
\caption{
Occupation probability of the wells for the elastic and inelastic
system at different values of $r = 1,0.9,0.8,0.7,0.6$, $M=64$ wells
and $N=512$ particles.  Elastic curve has a Poisson-like shape, given
by~\eqref{eq:nopoisson}, peaked around the average occupation $\lambda
= N/M = 8$. Curves referring to inelastic system deviate from the
elastic one. Inset: Entropic indicator $h$ as a function of the
inelasticity $r$.}
\label{fig:occup}
\end{figure}

In this case the distribution reads 
\begin{equation}
\Pi_{hr}[\{n_s\}] ={\cal N}\exp(-\Omega_{hr}[\{n_s\}]),
\label{eq:nopoisson}
\end{equation}
It displays an enhancement and a narrowing of the peak 
centered at $\lambda=N/M$,
because configurations characterized by highly 
populated compartments are unfavored (see Fig.~\ref{fig:occup}).
The narrowing of the peak can also be understood as a consequence
of the lower compressibility of the hard-rod system with respect to the
ideal one.

The presence of inelasticity introduces a different scenario with respect
to the elastic case. Indeed, the energy dissipation, strengthens
the correlations and contrasts the hard-core repulsion. 
Then, as the inelasticity parameter $r$ decreases,  
$P(n_i)$ deviates more and more from the elastic hard-rod distribution 
displaying a much larger variance, as shown in Fig.~\ref{fig:occup}.
The shape of $P(n_i)$ reflects the clustering tendency of the
inelastic system, because a slower decay of $P$ in the region $n_i>\lambda$
indicates the formation of clusters in few boxes. At the same time
the peak of the distribution shifts to lower values of $n_i$, showing
that the majority of the compartments are nearly empty: we
observe many wells containing just a 
small number of particles ($n_i\ll\lambda$) and few highly
populated wells, in contrast with the elastic case.

\begin{figure}[htbp]
\includegraphics[clip=true,keepaspectratio,width=8cm]{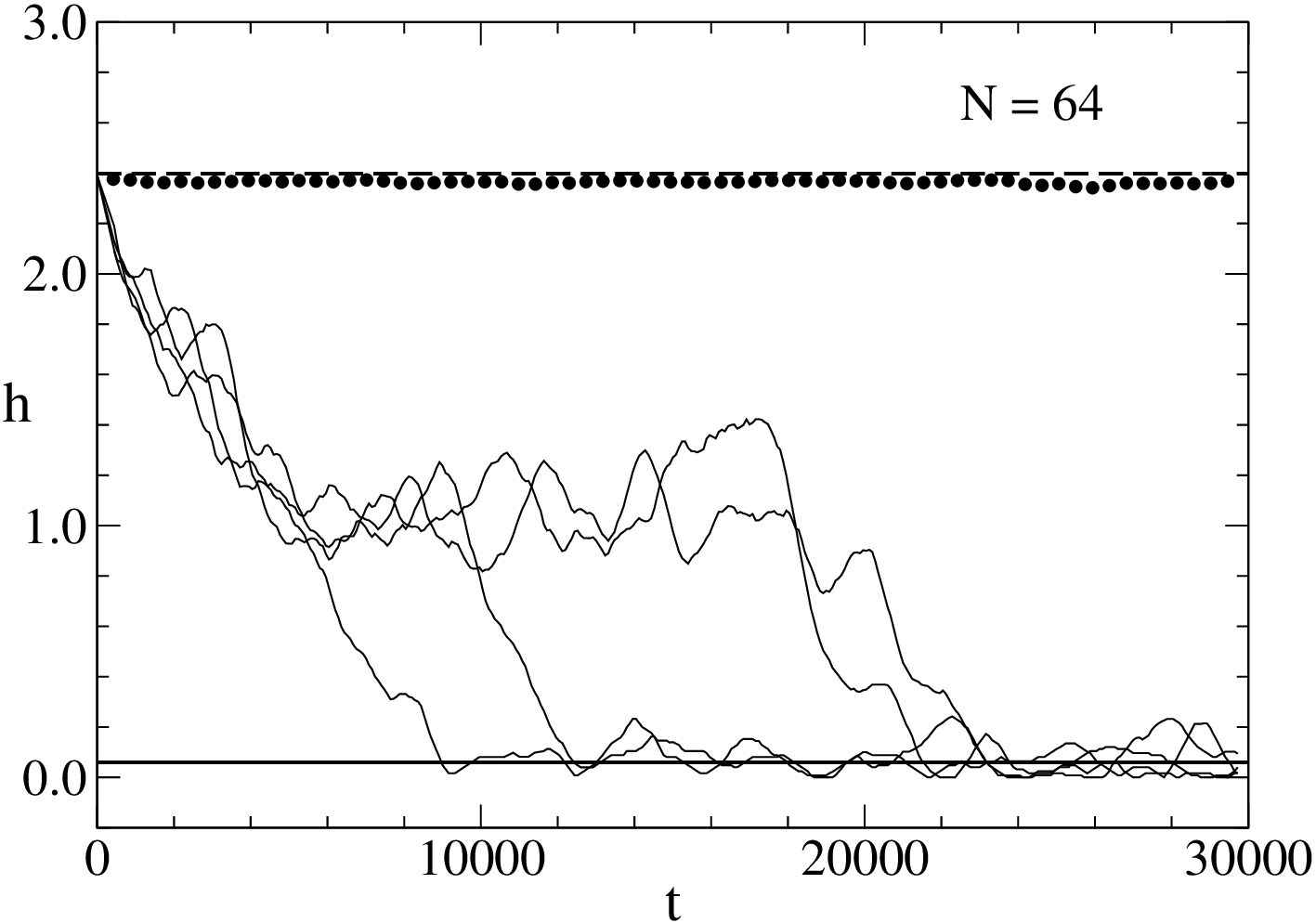}
\includegraphics[clip=true,keepaspectratio,width=8cm]{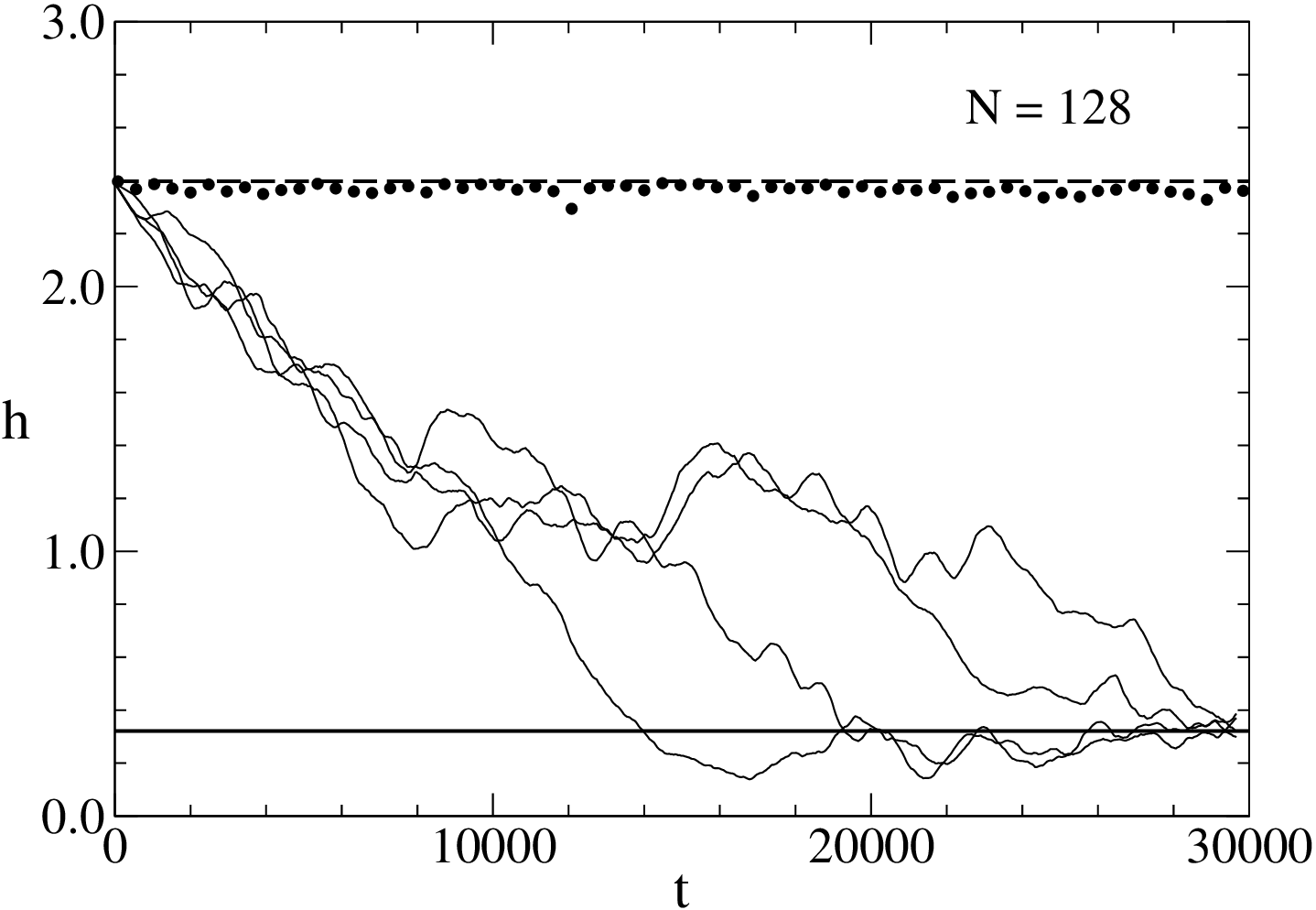}
\caption
{Evolution of the entropic parameter $h$, for a system starting from a 
uniform configuration (equi-populated wells), with 
$N=64$, $N=128$ particles of size $\sigma=0.1$, 
$M=11$ wells of width $15.0$ (density $=0.775$), parameter
$\alpha = 8.0$ and energy barriers $3.486$.  
The dashed reference line represents the uniform occupation $\ln(M)$,
the points correspond to the elastic system ($r=1$),
while the lower four curves refer 
to different initializations of velocities from a Maxwell distribution at 
temperature $T_b=2$, and inlesticity $r=0.8$. Thick horizontal line 
indicates the average of the asymptotic $h$-values over $40$ runs.}
\label{fig:entropy}
\end{figure}

The clustering phenomenon is well represented by the
following statistical indicator:
\begin{equation}
h=-\sum_{i=1}^{M} \frac{n_i}{N} \log \frac{n_i}{N}
\label{eq:entropia}
\end{equation}
The usefulness of $h$ in characterizing inhomogeneous particle
distribution in compartments is illustrated by considering the two
limiting cases.  When all particles pile up into a single well, $h \to
0$ (minimum value), whereas $h$ takes on its maximum value, $\ln(M)$,
when all the $M$ wells have identical populations. In intermediate cases,
the quantity $F=\exp(h)$ represents a measure of the average number of
occupied compartments. In the inset of Fig.~\ref{fig:occup} we report
the asymptotic value of $h$ as a function of $r$ referring to a system
of $N=512$ particles and $M=64$ compartments.   The value of $h$
decreases monotonically as the system becomes less elastic,
indicating that the grains tend to group together, reducing in this
way the fraction of occupied wells. Notice that as $r\to 1$ the
entropic indicator is slightly larger than the value, $h_R=4.0957$
concerning a Poisson distribution with the parameter tuned to the
simulation setup ($\lambda=N/M=8$). This discrepancy is a consequence
of excluded volume effects that result in the narrowing of the
distribution $P(n_i)$ discussed above (Eq,~\eqref{eq:nopoisson}).
The onset of clustering regimes can be emphasized by the time
evolution of $h$ and by making the direct comparison between the
elastic and inelastic cases (see Fig.~\ref{fig:entropy}). The system
is initially prepared in a state where the grains are non overlapping
and uniformly distributed among the $M = 10$ compartments.  Their
velocities are drawn according to a Maxwell distribution at the
temperature $T_b$ of the heat-bath. The dynamics drives the system
toward a new set of configurations. 

Figure~\ref{fig:entropy} shows that $h$ (dots)
for elastic particles fluctuates around a 
constant value
which is very close to the equi-populated limit $\ln(M)$ indicated
by the dashed line: as expected, there is no evidence of clustering
in the elastic system. 
On the other hand, the
clustering clearly appears from the curves of $h(t)$ referring to the 
inelastic system with $r=0.8$.   
After a transient regime, the indicator $h(t)$ becomes stationary, fluctuating
around a value which is considerably lower than the one corresponding 
to $r=1$.
This means that, due to clustering, only few wells result to be
effectively populated. 
The time at which the inelastic system departs from homogeneity depends 
crucially on the number of particles, as illustrated in 
Fig.~\ref{fig:entropy} at $N=64$ 
and $N=128$. 
We notice that the clustering cannot be considered complete because   
$h$ does not vanish but displays small fluctuations around a small constant
value. This implies that clusters exist as
dynamical structures, where particles can enter and exit. Otherwise, it may even happen that a cluster
already formed in a well breaks and reforms 
into another well after a long time.  
However, figure~\ref{fig:entropy} shows that the
observable $h(t)$ is a meaningful indicator, because it  
reaches a statistically steady value even when $\rho(x,t)$
is not fully stationary (as discussed in the previous section). 
We stress that if one performs an average of $\rho(x,t)$ on a 
time larger than the maximum escape time, the profile
$\overline{\rho(x,t)}$ would appear spatially uniform, while $h(t)$
for any time $t$ (after the initial transient) indicates that the
density is {\em non-uniform}. Therefore $h$ has to be considered the
correct order parameter to characterize density inhomogeneities in the
system.

\begin{figure}[htbp]
\includegraphics[clip=true,keepaspectratio,width=8.0cm]{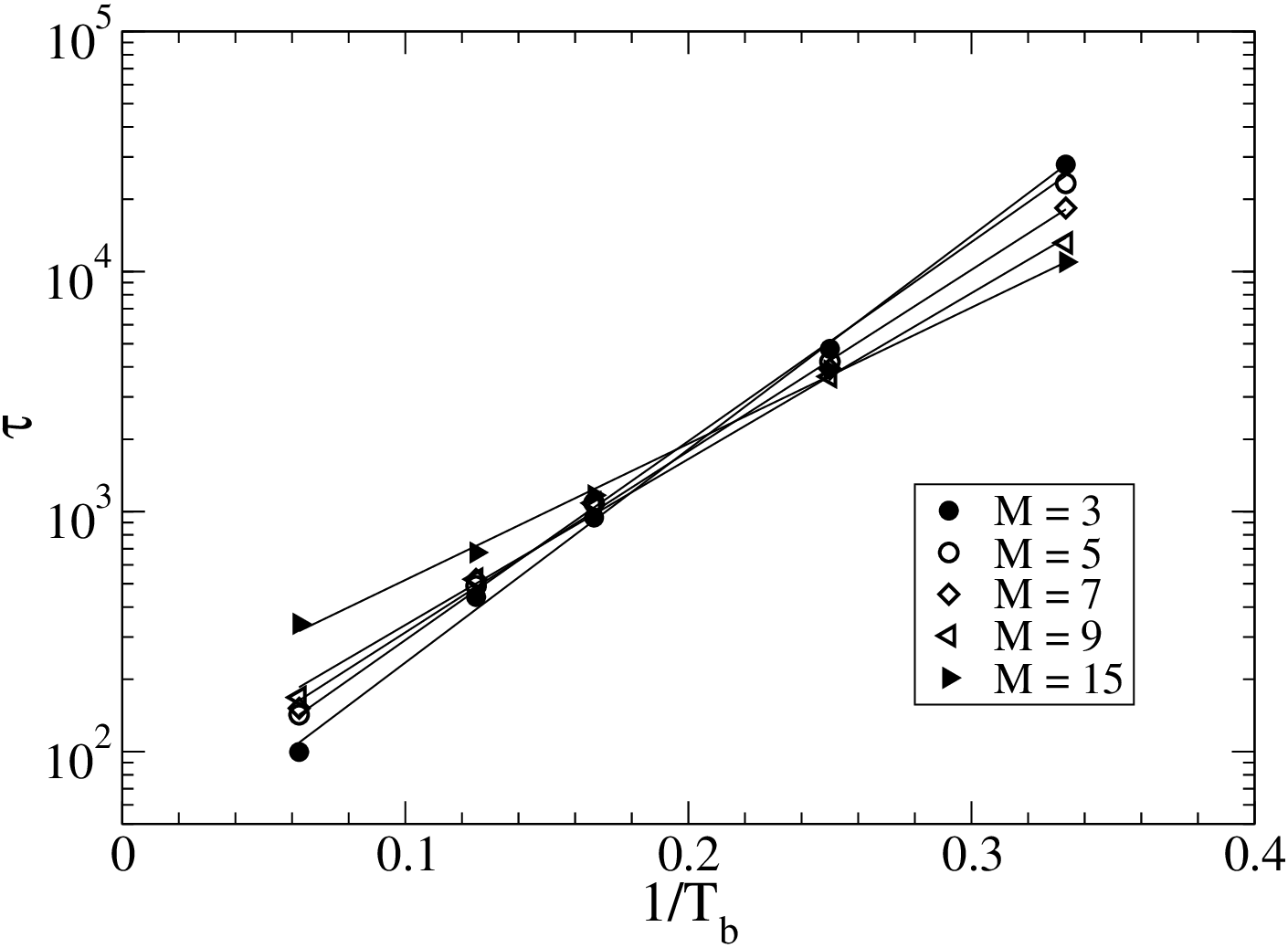}
\caption
{Average lifetime $\tau$ of a cluster initially placed in the central well
as a function of the heat-bath temperature $T_b$ ($\gamma=0.2$) 
at different number $M=3,5,7,9,15$ of wells, each of size $20.0$.
The average has been performed over a set of $100$ randomly selected
initial configurations in a granular system with $N=128$ particles 
of $\sigma=0.1$ 
and $r=0.7$, with $\alpha=8$ and potential barrier $\Delta V=3.486$}
\label{fig:esctime}
\end{figure}

We turn, now, our attention to a different process: we consider the
evolution of an initial clustered configuration, where all the $N$
particles at $t=0$ are located inside a central compartment in a
system made of $M=3,5,7,9,15$ wells.\cite{note}  We expect,
according to the level of the heat bath temperature $T_b$, two
different regimes: i) for large $T_b$ the cluster
decays toward the fully equidistributed state $n_i=N/M$ and
$\lim_{t\to \infty} F(t)=M$, since particles tend to fill neighboring
cells and distribute all over the system; ii) for low $T_b$ the
occupation of the compartment remains almost constant. It is useful
to define a characteristic time $t_s$, as the shortest time required
by the population of the central well, $n_0(t_s)$, to reach the fully
equidistributed value $\lambda=N/M$. Unlike the mean-field predictions,
\cite{Lohse0,Conti} such a lifetime is a strongly fluctuating
variable, depending on the initial configurations
and on the ``noise'' history. 
Therefore we computed its average $\tau =\langle t_s\rangle$
over $100$ different runs and different number of wells.
The results are reported in figure~\ref{fig:esctime} where the average 
lifetime $\tau$ is plotted
against the inverse of the heat bath temperature. We observe the
Arrhenius-like behavior~\cite{Biwell} with slopes changing with the
total number of boxes $M$, in  qualitative agreement with the
analytical predictions.\cite{Lohse0,Conti}  
The larger the number of wells, the smaller the
slope, indicating that the most stable cluster is the one with only
three wells. We also remark that we were not able to observe a sudden
death of the cluster as predicted by theoretical approaches.

Among the theoretical treatments of compartmentalized granular gases,
the flux-model\cite{Eggers,Lohse0,Lohse1,Lohse2,Droz} enjoys a great 
popularity in
view of its simplicity. At the heart of this approach is the so called
flux-function $\Phi(n)$, which represents a measure of the number of
particles crossing a section of the system per unit time. The form of
the flux-function is admittedly phenomenological and follows from the
experimental observations. The latter require $\Phi(n)$ to be, at
fixed driving intensity, an increasing function for moderate values of
the occupation number, $n$, and a decreasing function for larger
values of $n$, to account for the stronger dissipation at high
densities.  Such a non monotonicity entails the possibility of
spatially non uniform steady solutions.\cite{Lohse0}

\begin{figure}[htbp]
\includegraphics[clip=true,keepaspectratio,width=8.0cm]{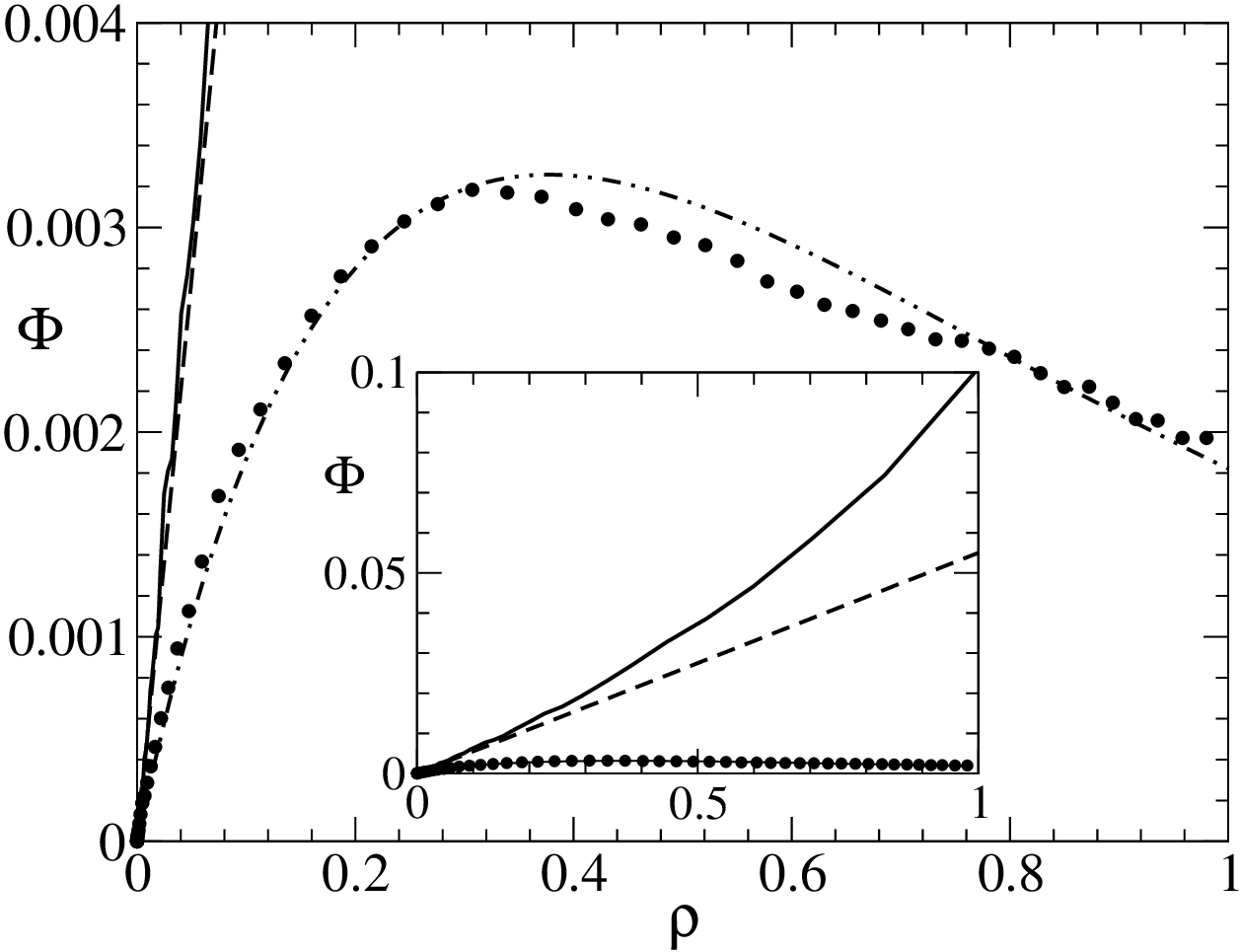}
\caption
{Flux versus density for a single potential well (M=1) with absorbing 
boundaries. The curves refer to independent (dashed), elastic (full) and 
inelastic (circles) particles. 
Simulations parameters are: initial occupation number  
$N=50$, $\sigma=0.1$, $\alpha=8$, well size $w = 50$ 
(barrier $\Delta V= 3.486$), 
bath temperature $T_b=1.0$ and coefficient of restitution $r=0.8$.
The theoretical prediction of the flux described in the appendix
is the dot-dashed plot.   
The inset reports the same curves on a larger scale.}
\label{fig:flux}
\end{figure}

For the sake of comparison, we extract from simulations the
flux function for our model, by measuring the number density of particles, $\rho$,
flowing outside a given well per unit time as a function of the density
of particles within that well. The simulations start from an
initial configuration where all the $N_0=50$ particles are located in the 
same well of width $w=25$ ($\rho_0=N_0/w$) and height $\Delta V = 3.486$.  
Once the particles reach the top of
the barrier, they are not allowed to re-enter (absorbing boundary
conditions). As time goes on, we record the
density, $\rho_{out}(t)$, of particles escaping from the well and
average it over an ensemble of $1000$ randomly selected initial
configurations. The averaging procedure yields a curve 
$\langle \rho_{out}(t)\rangle$ from which we estimate its time 
derivative numerically. Finally, a parametric plot of 
$d \langle \rho_{out}(t)\rangle/dt$  
versus the density of the remaining particles  in the well, 
$\rho_0 -\langle \rho_{out}(t)\rangle$, 
gives the flux function $\Phi(\rho)$ of the model. 
It is instructive to compare the properties of the flux
relative to three different systems: non interacting, elastically colliding and 
inelastically colliding particles, as shown in figure~\ref{fig:flux}. 
In the low density limit the interactions are almost negligible 
and the flux functions display a linear behaviors characterized by a common slope. 
The situation becomes more interesting when the well
contains more particles.  The flux for non-interacting particles  
grows linearly with $\rho$ (dashed curve): the larger $\rho$, the larger is the
probability to observe an escape event. While, in the system of elastic hard
particles, the hard-core potential produces two competing effects:
on one hand the repulsion makes the escape rate for particles near the
border of the well larger than the corresponding
rate for non-interacting particles. On the
other hand, the same repulsion induces a cage effect entrapping 
the inner particles and tends to reduce the escape rate. However,  from
the dot-dashed curve in the inset of figure~\ref{fig:flux}, we see that the first effect overcomes the
second enhancing the flux with respect to independent particles.   
Finally, in the inelastic case, 
the tendency of the  particles to form clusters and to dissipate
kinetic energy greatly decreases the exit probability from the well,
so that the flux is much smaller (almost two orders of magnitude) than the flux for the independent and elastic 
systems. We observe a slow increase with $\rho$ and a local maximum clearly visible in dot curve of  Fig.~\ref{fig:flux}. 
The dot-dashed line, in the same figure, represents instead the theoretical 
attempt to explain the behaviour of $\Phi(\rho)$  
through a Kramers' theory which combines the excluded volume effect and the 
inelasticity into the escape-time formula~(\ref{newkramer}) (see Appendix). 
The theoretical curve is plotted for all the same parameters of the 
simulations except for a unknown prefactor which is used as a fitting 
coefficient.

\section{Conclusions}
In this paper, inspired by a series of recent experiments, we have 
introduced and studied a model of randomly driven inelastic
hard rods in the presence of an external spatially periodic
potential. The one dimensional model
presents some of the salient
features of higher dimensional models, such as packing, clustering,
velocity correlations.\cite{CDBP}
By comparison with the elastic hard rod model, used here as reference
system,
we found that the inelasticity has deep repercussions on the
configurational properties.
The fluctuations in density result increased and show the tendency of
the system to cluster. Moreover, the (granular) temperature profile
ceases to be spatially uniform showing minima in correspondence of
the density maxima. As a measure the density fluctuations
we have introduced an ``entropic'' indicator, $h$, which provides
information about the statistics of the partition of the grains in the
various wells. We also performed  numerical experiments concerning
the stability of a cluster obtained by  initially concentrating
all the particles in a single well.
Finally, we considered within the model the shape of the flux function
which plays a major role in the macroscopic description of
compartmentalized systems.

\acknowledgments We thank Massimo Cencini for very useful discussions 
and suggestions. U.M.B.M. acknowledges the financial support by 
COFIN-MIUR 2003020230 and F.C. acknowledges the financial support by 
FIRB-MIUR RBAU013LSE-001.

\appendix*\section{Flux Function}
\begin{figure}[htbp]
\includegraphics[clip=true,keepaspectratio,width=8.0cm]{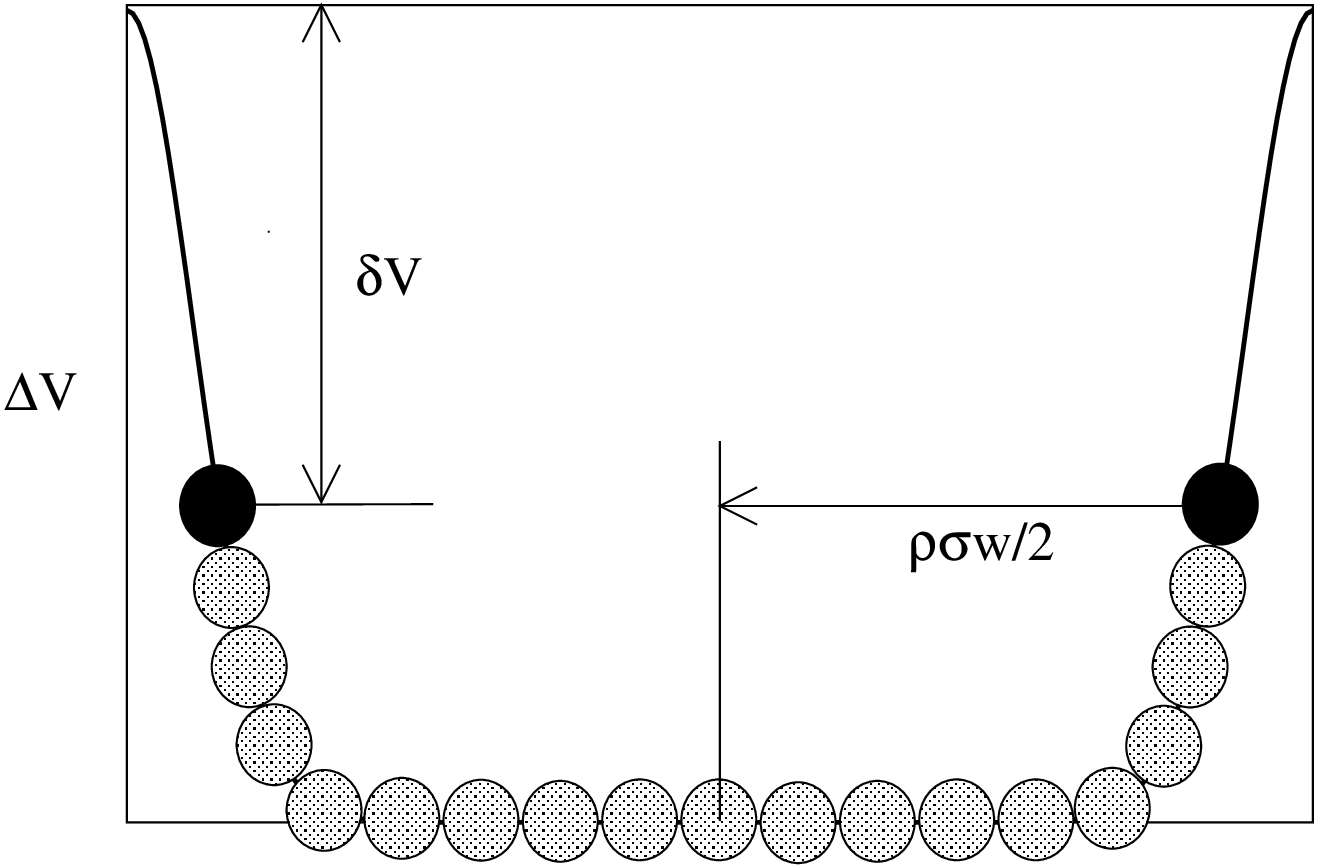}
\caption
{Sketch of the reduction in the barrier height due to the excluded 
volume effect and leading to the renormalized flux function~(\ref{pippo})}
\label{fig:example}
\end{figure}

In this appendix we present a theoretical argument to explain the behaviour 
of flux function $\Phi$ measured in simulations.  
The flux of particles outside of a potential well, with size $w$, height $\Delta V$  and 
populated with a density $\rho$, can be computed via the formula
\begin{equation}
\frac{d\rho}{dt} = \Phi(\rho) = \frac{\rho}{\tau(\rho)}\;
\label{eq:flux_th}
\end{equation}
where $\tau(\rho)$ is the particle escape time across the barrier. For the 
present granular system, 
this time follows the Kramers-Arrhenius law for independent particles,  
but with parameters renormalized by interactions. As already   
mentioned, $\tau(\rho)$ is the outcome of two competing effects, the excluded 
volume and the granular dissipation. The dissipation determines the 
granular temperature that can be approximated by the solution of Eq.~(\ref{eq:Tg}), 
which holds only for a inelastic system without external potential. 
However for the potential parameters chosen in our 
simulations, formula ~(\ref{eq:Tg}) can be considered a fair approximation 
of the temperature in a well, because wells are rather flat and barriers so steep that 
particles feel the confining effect of the 
potential only in the neighborhood of well boundaries (Fig.\ref{fig:example}). 
The excluded volume interaction, instead, 
is responsible for a mean reduction of the potential barrier $\Delta V$.  
This effect is pictorially illustrated in figure \ref{fig:example}, 
where the leftmost/rightmost particles (darker), the only ones allowed to jump, 
are displaced from the well center $x_c$ of an amount 
$\rho\sigma w/2$. They experience an effective energy-barrier 
roughly estimated as
\begin{equation}
\delta V (\rho) = V(x_c + w/2) - 
V(x_c + \rho\sigma w/2)
\label{pippo}
\end{equation}
and the renormalized Arrhenius law becomes     
\begin{equation}
\tau(\rho) = \frac{1}{\nu} \exp\{\delta V(\rho)/T_g(\rho)\}
\end{equation}
Simulations indicate a very weak dependence of $\nu$ on $\rho$ 
for the parameters we employed. Therefore, the prefactor $\nu^{-1}$ has
been considered as a adjustable time scale.
Numerical solutions of equation~(\ref{eq:Tg}) computed at different 
densities allow to construct the curve of the flux $\Phi$ as a function of 
$\rho$, reported in figure \ref{fig:flux} in dot-dashed style.

\begin{thebibliography}{99}
\bibitem{note} 
We properly shift the potential to bring the minimum of the central well
to coincide with the origin of $x$-axis: $V(x) \to V(x-w/2)$.
\end{thebibliography}
\end{document}